# MODELING ASTRONOMICALLY OBSERVED INTERSTELLAR INFRARED SPECTRA BY IONIZED CARBON PENTAGON-HEXAGON MOLECULES $(C_9H_7)^{n+}$

NORIO OTA

Graduate School of Pure and Applied Sciences, University of Tsukuba,
1-1-1 Tenoudai Tsukuba-city 305-8571, Japan; n-otajitaku@nifty.com

Modeling a promising carrier of the astronomically observed polycyclic aromatic hydrocarbon (PAH), infrared (IR) spectra of ionized molecules $(C_9H_7)^{n+}$ were calculated based on density functional theory (DFT). In a previous study, it was found that void induced coronene $C_{23}H_{12}^{++}$ could reproduce observed spectra from 3 to 15μm, which has carbon two pentagons connected with five hexagons. In this paper, we tried to test the simplest model, that is, one pentagon connected with one hexagon, which is indene like molecule $(C_9H_7)^{n+}$ (n=0 to 4). DFT based harmonic frequency analysis resulted that observed spectrum could be almost reproduced by a suitable sum of ionized $C_9H_7^{n+}$ molecules. Typical example is $C_9H_7^{++}$. Calculated peaks were 3.2, 7.4, 7.6, 8.4, and 12.7μm, whereas observed one 3.3, 7.6, 7.8, 8.6 and 12.7μm. By a combination of different degree of ionized molecules, we can expect to reproduce total spectrum. For a comparison, hexagon-hexagon molecule naphthalene $(C_{10}H_8)^{n+}$ was studied. Unfortunately, ionized naphthalene shows little coincidence with observed one. Carbon pentagon-hexagon molecules may play an important role as interstellar molecular dust.

Key words: astrochemistry - infrared: numerical - molecular data: PAH – indene - naphthalene:

## 1, INTRODUCTION

Interstellar dust shows ubiquitous specific infrared (IR) spectrum from 3 to 20μm (Boersma et al. 2014). Recently, Tielens (Tielens 2013) discussed that void induced polycyclic aromatic hydrocarbons (PAH's) may be one candidate. In previous papers (Ota 2014, 2015a, 2015b), infrared calculations on void coronene $C_{23}H_{12}^{++}$ was tried to test this expectation. Result was amazing that this single molecule could almost reproduce a similar IR spectrum with astronomically well observed one. One example is illustrated in Figure 1, where upper figure was astronomically observed four sources spectra edited by (Boersma 2009), whereas red curve in a lower figure show calculated one (Ota 2014) based on density functional theory (DFT). Broken lines are well observed ubiquitous wavelength. We can see a fairly good coincidence with each other.

Current central concept to understand the observed astronomical spectra is the decomposition method from the data base of many PAHs experimental and theoretical analysis (Boersma et al. 2013, 2014). Question is why previous calculation of single molecule $C_{23}H_{12}^{++}$ could almost reproduce observed spectra. Remarkable feature is a molecular structure as shown in left side of Figure 1, that is, carbon two pentagons were connected with five hexagons. This triggered one idea that the simplest pentagon-hexagon molecule may be able to explain fundamental IR behavior. Indene like molecule $(C_9H_7)$ is a typical example as shown in Figure 2 (b). This paper opens IR spectra calculation of ionized indene like molecules $(C_9H_7)^{n+}$ (n=0 to 4) compared with that of a hexagon-hexagon combined naphthalene $(C_{10}H_8)^{n+}$. A capability of reproducing observed spectra is demonstrated by ionized $C_9H_7$ molecules.

## 2, MODEL MOLECULES

As illustrated in Figure 2(a), starting molecule is naphthalene $(C_{10}H_8)$, which has carbon two hexagons showing typical PAH characteristics. In interstellar space, especially in new star born area, many high energy protons and high energy photon may attack naphthalene. Proton $H^+$ attacks one particular carbon site of $C_{10}H_8$ and creates a single void. After that, recombination of molecule may bring to $C_9H_7$, which is shown in Figure 2(b). High energy photon also attacks such molecules to ionize deeply such as $C_9H_7, C_9H_7^+, C_9H_7^{++}$ and so on $(C_9H_7)^{n+}$ (n=0 to 4). For every ionized state, after atom position optimization, IR spectrum calculation should be done. For comparison, IR spectrum of ionized naphthalene $(C_{10}H_8)^{n+}$ was also obtained.

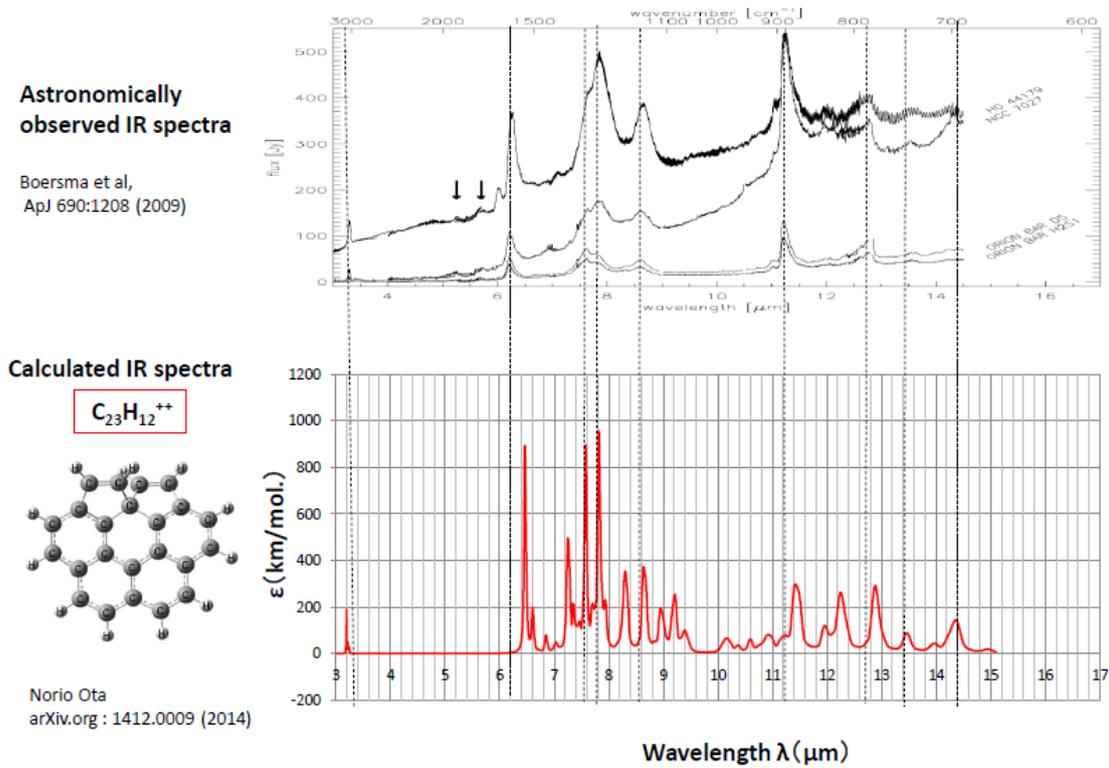

**Figure 1** Upper figure is astronomically observed infrared spectra (Boersma 2009) for four souses. Lower one is calculated one for void coronene $C_{23}H_{12}^{++}$ (Ota 2014). Single molecule spectrum could almost reproduce observed one.

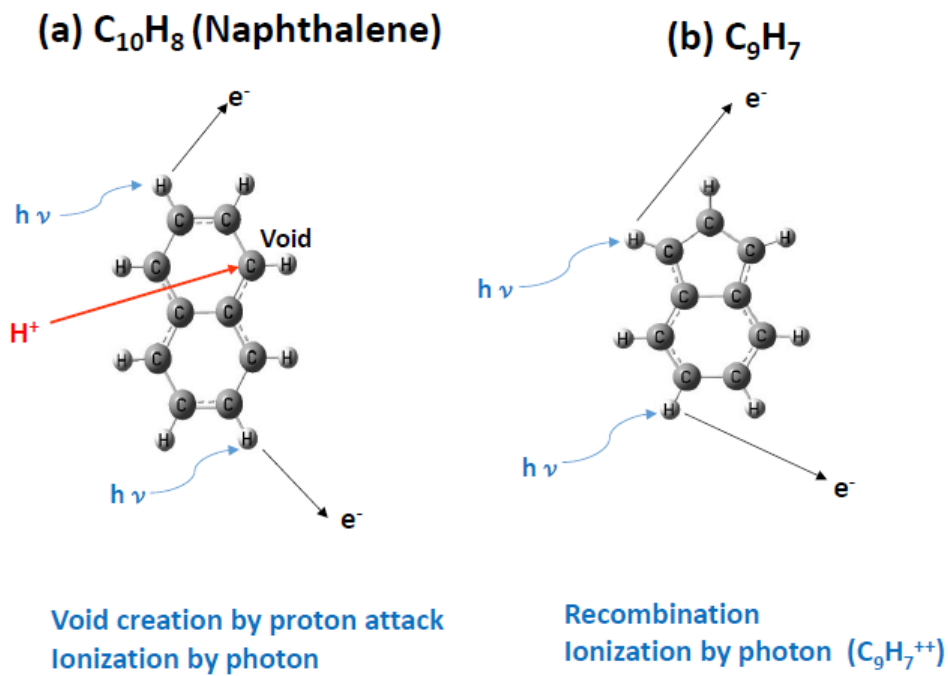

**Figure 2** Modeling a creation of void by proton attack on naphthalene and molecular recombination brings an indene like molecule $C_9H_7$. High energy photon irradiation causes ionization for every molecule.

## 3, CALCULATION METHOD

We have to obtain total energy, optimized atom configuration, and infrared vibrational mode frequency and strength depend on a given initial atomic configuration, charge and spin state Sz. Density functional theory (DFT) with unrestricted B3LYP functional (Becke 1993) was applied utilizing Gaussian09 package (Frisch et al. 2009, 1984) employing an atomic orbital 6-31G basis set. The first step calculation is to obtain the self-consistent energy, optimized atomic configuration and spin density. Required convergence on the root mean square density matrix was less than $10^{-8}$ within 128 cycles. Based on such optimized results, harmonic vibrational frequency and strength was calculated. Vibration strength is obtained as molar absorption coefficient ε (km/mol.). Comparing DFT harmonic wavenumber $N_{DFT}$ (cm$^{-1}$) with experimental data, a single scale factor 0.965 was used (Ota 2015b). For the anharmonic correction, a redshift of 15cm$^{-1}$ was applied (Ricca et al. 2012).

Corrected wave number N is obtained simply by N (cm$^{-1}$) = $N_{DFT}$ x 0.965 – 15 .

Also, wavelength λ is obtained by λ (μm) = 10000/N(cm$^{-1}$).

## 4, IR SPECTRA OF IONIZED INDENE LIKE MOLECULE $(C_9H_7)^{n+}$

In Figure 3, vertical value show integrated absorption coefficient ε (km/mol), whereas horizontal line shows wavelength λ(μm). Blue curve was calculated Lorentzian type distribution with the full width at half-maximum (FWHM) of 15 cm$^{-1}$ based on the harmonic intensity. Neutral $C_9H_7$ is radical and has a spin parameter Sz=1/2 as noted in Table 1, which is unstable in conventional circumstance. However, in interstellar space, material density is so small (less than 1 molecule in 1cm$^3$), survived life time will be long sufficient to be observed. There are two main peaks as shown in Figure 3(a), which are 3.2 and 13.5μm, which values are close to observed 3.3 and 13.4μm. Mono-cation $C_9H_7^+$ has fairly large electric dipole moment of 1.00 Debey (see Table 1). Calculated IR spectrum became complex as illustrated in (b). Major peaks were 6.2, 7.4, 7.7, 8.6, and 12.8μm which were related to observed 7.6, 7.8, 8.6, and 12.7μm. Peak height of 8.6 and 12.8μm were fairly large. Not coincided peaks were 6.8 and 7.0μm. Similar tendency was obtained in case of di-cation $C_9H_7^{++}$ as shown in (c). Peaks at 3.2, 7.4, 7.6, 8.4, and12.7μm were close to observe one. Unfortunately, peaks at 7.0 and 10.5μm were not fit with observed one. In tri-cation $C_9H_7^{+++}$ case in (d), there are five fairly nice peaks of 3.3, 7.4, 7.7, 8.4, and 11.2μm. Also, quadri-cation $C_9H_7^{++++}$ in (e) shows four nice peaks as summarized in Table 1.

Table 1 is useful to survey total behavior of ionized $C_9H_7$ molecules. Map of green colored number (calculated wavelength) could totally reproduce astronomically well observed wavelength marked by orange in left column. We could imagine that in interstellar space indene-like molecules may be floating as different degree of ionized states. Ionization energy is important to realize such coexistence of different ionized states. We can compare ionized energy in third line of Table 1. For mono-cation creation we need 7.8eV, and in order to obtain more deep ionized state, we need 20~60eV or more. There are three capabilities, one is multiple photon absorption at the same time, next is photon irradiation from ultrahigh temperature new born star, and other capability is direct proton and/or charged carbon collision with molecules. We need laboratory experiments for confirmation.

## 5, IONIZED NAPHTHALENE $(C_{10}H_8)^{n+}$

Ionized naphthalene $C_{10}H_8$ is interesting to compare with ionized $C_9H_7$. Neutral naphthalene is stable and typical example of PAH. Electric dipole moment is completely zero and spin parameter Sz is also zero to be non-magnetic. Calculated IR was shown in Figure 4(f). We can see only two major peaks at 3.2 and 12.9μm, which correspond to observed 3.3 and 12.7μm. Other ionized naphthalene $(C_{10}H_8)^{n+}$ (n=1 to 3) show little coincidence with observed one. As an example, calculated IR of $C_{10}H_8^{+++}$ was illustrated in Figure 4(g). Only two wavelength 3.2 and 11.2μm match with expected one. However, other peaks, 7.1, 7.2, 7.4, 10.2, 13.8μm did not coincided with observed peaks. Major peaks were summarized in Table 2. Number of green colored number (coincided frequency) is few and black color (not coincided) are many. Comparing with ionized $C_9H_7$, we cannot expect to cover total observed spectra by naphthalene group. We need carbon pentagon-hexagon molecular structure to reproduce interstellar IR spectra.

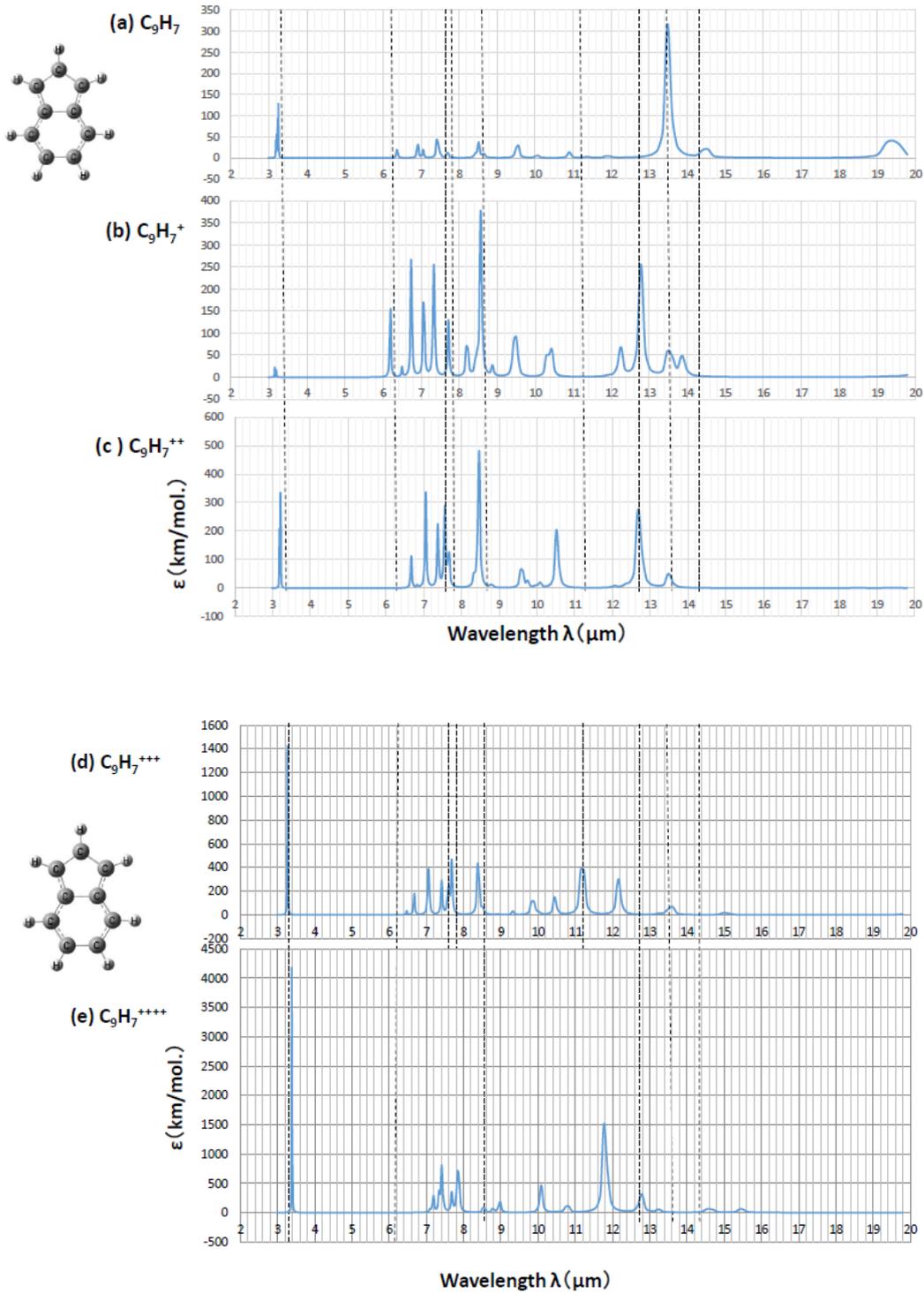

**Figure 3** Calculated infrared absorption spectrum of (a) $C_9H_7$, (b) $C_9H_7^+$, (c) $C_9H_7^{++}$, (d) $C_9H_7^{+++}$, and (e) $C_9H_7^{++++}$. Vertical broken lines show astronomically well observed wavelength.

Table 1 Calculated fundamental properties and major infrared absorption wavelength of ionized $C_9H_7$. Green colored numbers are coincided wavelength with observed one, whereas black one are not.

**$C_9H_7$**

| Charge | 0 (e) | +1 | +2 | +3 | +4 |
|---|---|---|---|---|---|
| Spin | Sz=1/2 | 0 | 1/2 | 0 | 1/2 |
| Energy Difference | 0 (eV) | 7.3 | 20.5 | 39.6 | 65.3 |
| Dipole moment (Debey) | 0.00 | 1.00 | 0.68 | 0.92 | 1.06 |
| Symmetry For calculation | C1 | C1 | C1 | C1 | C1 |
| Observed Wavelength (μm) | Calculated major wavelength (μm) | | | | |
| 3.3 | 3.2 | | 3.2 | 3.3 | 3.4 |
| 6.2 | | 6.2 | | | |
|  | | 6.8, 7.0 | 7.0 | 7.1 | |
| 7.6 | | 7.4 | 7.4 | 7.4 | 7.4 |
| 7.8 | | 7.7 | 7.6 | 7.7 | 7.8 |
| 8.6 | | 8.6 | 8.4 | 8.4 | |
|  | | | 10.5 | | 10.1 |
| 11.2 | | | | 11.2 | |
|  | | | | | 11.8 |
| 12.7 | | 12.8 | 12.7 | | 12.8 |
| 13.4 | 13.5 | | | | |
| 14.5 | | | | | |

Table 2 Calculated fundamental properties and major infrared absorption wavelength of ionized naphthalene $C_{10}H_8$. Poor coincidence with observed one.

**$C_{10}H_8$ (Naphthalene)**

| Charge | 0 (e) | +1 | +2 | +3 | +4 |
|---|---|---|---|---|---|
| Spin | Sz=0/2 | 1/2 | 0/2 | 1/2 | 0/2 |
| Energy | 0 (eV) | 7.8 | 20.8 | 39.6 | 64.1 |
| Dipole moment (Debey) | 0.00 | 0.00 | 0.00 | 0.00 | 0.00 |
| Symmetry | C1 | C1 | CI | C2H | D2H |
| Observed Wavelength (μm) | Calculated major wavelength (μm) | | | | |
| 3.3 | 3.2 | | 3.2 | 3.2 | 3.3 |
| 6.2 | | | | | |
|  | | 6.7 | 6.9, 7.2 | 7.2 | |
| 7.6 | | | | | 7.4 |
| 7.8 | | | | | |
|  | | 8.3 | 8.2 | | |
| 8.6 | | | | | |
|  | | | 9.6 | 10.2 | |
| 11.2 | | | | 11.2 | |
|  | | | | | 12.2 |
| 12.7 | 12.9 | | | | |
| 13.4 | | 13.2 | | | |
|  | | | 13.7 | 13.9 | |
| 14.5 | | | | | |

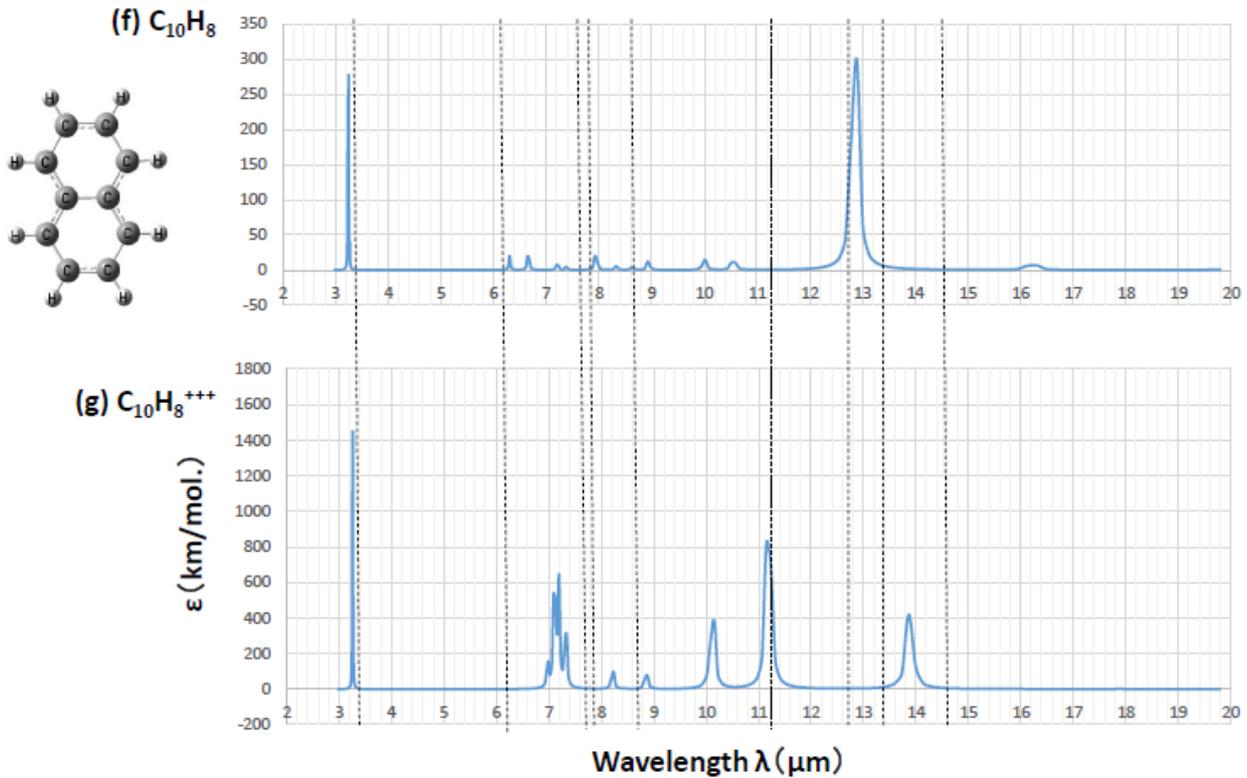

**Figure 4** Calculated infrared absorption spectrum of (f) $C_{10}H_8$ and (g) $C_{10}H_8^{+++}$

### 6, CONCLUSION

Modeling a promising carrier of the astronomically observed polycyclic aromatic hydrocarbon (PAH), IR spectra of ionized molecules $(C_9H_7)^{n+}$ were calculated based on DFT based Gaussian program.

(1) Void induced di-cation coronene $C_{23}H_{12}^{++}$ could reproduce astronomically well observed spectra from 3 to 20µm. Molecular configuration is characteristic that carbon two pentagons directly connected with five hexagons.
(2) In order to understand IR characteristics, we tried to test the simplest model, that is, one pentagon connected with one hexagon, which is indene like molecule $(C_9H_7)^{n+}$. Density functional theory based harmonic frequency analysis was done.
(3) Well observed astronomical infrared spectrum from 3-15µm could be almost reproduced by a series of ionized molecules $(C_9H_7)^{n+}$ (n=0 to 4). In case of di-cation $C_9H_7^{++}$, calculated peaks were 3.2, 7.4, 7.6, 8.4, and 12.7µm which correlated with observed 3.3, 7.6, 7.8, 8.6 and 12.7µm.
(4) By a combination of different degree of ionized molecules, we can believe to reproduce total observed spectrum.
(5) For a comparison, ionized naphthalene $(C_{10}H_8)^{n+}$ was studied. Unfortunately, ionized naphthalene show little coincidence with observed on.
(6) Carbon pentagon and hexagon coupled molecules may play an important role as interstellar molecules.

ACKNOWLEDGEMENT

I would like to say great thanks to Dr. Christiaan Boersma, NASA Ames Research Center, to permit me to refer a figure (Boersma et al. 2009), also thanks his kind and useful suggestions.

REFERENCES

Detailed references concerning infrared observation, laboratory experiment, and theoretical analysis were noted by Ricca et al (Ricca 2012). Excellent review on molecular universe was opened by Prof. A. Tielens (Tielens 2013).

Submitted to arXiv.org  :  **June** 18, 2015 by Norio Ota